\documentclass[11pt,a4paper]{article}
\pdfoutput=1
\usepackage{jheppub}

\usepackage{epsfig}
\usepackage{amssymb}
\usepackage{amsbsy}
\usepackage{amsmath}
\usepackage{url}

\newcommand{\rf}[1]{(\ref{#1})}
\newcommand{\beq}{\begin{equation}}
\newcommand{\eeq}{\end{equation}}
\newcommand{\be}{\begin{equation}}
\newcommand{\ee}{\end{equation}}
\newcommand{\bea}{\begin{eqnarray}}
\newcommand{\eea}{\end{eqnarray}}
\newcommand{\eq}[1]{Eq.~(\ref{#1})}
\newcommand{\non}{\nonumber \\*}
\newcommand{\ie}{{i.e.}\ }
\newcommand{\bv}{\begin{verbatim}}
\newcommand{\ev}{\end{verbatim}}

\newcommand{\vp}{\varphi}
\newcommand{\bvp}{{\boldsymbol \varphi}}
\newcommand{\dr}{{\delta \rho}}
\newcommand{\e}{\,\mbox{e}}
\renewcommand{\d}{{\rm d}}

\newcommand{\blambda}{\bar\lambda}
\newcommand{\brho}{\bar\rho}
\newcommand{\C}{\blambda}

\newcommand{\half}{{\textstyle \frac 12}}

\newcommand{\tr}{\mathrm{tr}}
\newcommand{\LA}{\left\langle}
\newcommand{\RA}{\right\rangle}

\def\gtrsim{\mathrel{\mathpalette\fun >}}
\def\fun#1#2{\lower3.6pt\vbox{\baselineskip0pt\lineskip.9pt
\ialign{$\mathsurround=0pt#1\hfil##\hfil$\crcr#2\crcr\sim\crcr}}}

\def\ga{\gtrsim} 
\begin{document}


\title{Mean field quantization of effective string}

\author{Yuri Makeenko}

\affiliation{Institute of Theoretical and Experimental Physics,\\
B. Cheremushkinskaya 25, 117218 Moscow, Russia}

\emailAdd{makeenko@itep.ru}


\abstract{I describe the recently proposed quantization 
of bosonic string about the mean-field ground state, paying 
special attention to the differences from the usual quantization
about the classical vacuum which turns out to be unstable for $d>2$.
In particular, the string susceptibility index $\gamma_{\rm str}$ is
1 in the usual perturbation theory, but
equals 1/2 in the mean-field approximation that applies for $2<d<26$.
I show that the total central charge equals zero in the mean-field approximation
and argue that  fluctuations about the mean field do not spoil conformal invariance.}



\maketitle

\section{Introduction}

Strings or more generally two-dimensional random surfaces have wide applications in physics:
from biological membranes to QCD. However, a nonperturbative theory of quantum strings,
which goes back to 1980's, makes sense only if the dimension of target space $d\!<\!2$,
where the results both of dynamical triangulation \cite{KKM85,Dav85,ADF85} and of conformal field 
theory~\cite{KPZ,Dav88,DK89} are consistent and agree. For $d>2$
the scaling limit of dynamically triangulated random surfaces
is particle-like\footnote{A detailed description can be found in the book~\cite{ADJ97}.}
rather than  string-like because only the lowest mass scales but the string tension does not 
scale and tends to infinity in the scaling limit.\cite{AD87}
Analogously, the conformal field 
theory approach does not lead to sensible results for $2<d<26$.\cite{KPZ,Dav88,DK89}
The conclusion was that quantum string 
does not exist nonperturbatively for $2<d<26$, while it beautifully works for $d<2$.
However, we understand that strings do exist as physical objects
 in $d=4$ space-time dimensions. 

A potential way out was to adopt the viewpoint that string is not a fundamental 
object but is rather formed by more fundamental degrees of freedom.
This philosophy perfectly applies to QCD string,%
\footnote{For a brief introduction see e.g.~\cite{Mak12}.}
where these are  fluxes of the gauge field.
String description  makes sense only  for the distances larger than the confinement scale.
For shorter distances the quark-gluon degrees of freedom are more
relevant due to asymptotic freedom. This picture is well justified both
by experiment and by lattice simulations. The string tachyon which is
a short-distance phenomenon does not show up in the QCD spectrum.

A breakthrough along this line is due to the ``effective string'' philosophy~\cite{PS91},
which works perturbatively order by order in the inverse string length
(for recent advances see~\cite{DFG12,AK13,Hel14,Bra16}).
Then string quantization is consistent even below 
 the critical dimension ($d=26$ for the
relativistic bosonic string) and a few leading orders 
reproduce~\cite{PS91,Dru04,AFK11} 
the Alvarez-Arvis ground-state energy~\cite{Alv81,Arv83,Ole85}.
In this Paper I shall pay much attention to this issue.

In the recent series of papers~\cite{AM15,AM17a,AM17c} it has been understood why
lattice string formulations resulted in the particle-like continuum limit.
A nonperturbative mean-field solution of the Nambu-Goto string showed that
the usual classical vacuum about which string is quantized is unstable for $d>2$,
while another nonperturbative vacuum is stable for $2<d<26$, like it happens in the well-known example
of the two-dimensional $O(N)$ sigma-model.
For the true ground state the value of 
the metric at the string worldsheet ($\rho_{ab}= \brho \delta_{ab}$ in the 
conformal gauge) becames infinite in the scaling limit.
For this reason an infinite amount of stringy modes (which is $\propto \brho /a^2$ with
$a$ being a UV cutoff)  can be reached
even at the distances of order $a$, That was in contrast to the usual  continuum limit in quantum field theory, where 
the amount of degrees of freedom can be infinite only if the correlation length is infinite.
The discovered phenomenon is specific to theories with diffeomorphism  invariance and
was called the {\em Lilliputian}\/ continuum limit.

The task of this Paper is to analyze properties of the mean-field vacuum which
plays the role of a ``classical'' string ground state and
``quantum'' fluctuations about it. I put here quotes to emphasize this state is 
a genuine nonperturbative quantum state from the viewpoint of the usual 
semiclassical expansion in $\alpha'$ about the classical vacuum. 
We thus perform a resummation of this expansion with the leading order given
by the sum of bubble-like diagrams.
An analogy with the two-dimensional $O(N)$ sigma-model at large $N$ can be
again instructive. 
I shall pay special attention to a comparison with the 
Knizhnik-Polyakov-Zamolodchikov (KPZ) -- David-Distler-Kawai (DDK) results~\cite{KPZ,Dav88,DK89}
 for the parturbative vacuum, which is applicable for $d<2$.

It will be shown in the Paper that the total central charge of the system vanishes in the 
mean-field approximation which is thus consistent in noncritical dimension \mbox{$2<d<26$}.
This is in contrast to  the old canonical quantization which works only in
the critical dimension $d=26$ and compliments the effective-string approach
of Polchinski-Strominger, where the consistency is explicitly
demonstrated to a few lower orders
of the perturbative expansion~\cite{PS91,Dru04,AFK11}.
I then analyze a ``semiclassical'' correction to the mean-field approximation
and show  that it does not spoil conformal invariance in spite of 
logarithmic infrared divergences caused by the propagator of a massless field,
which cancel in the sum of diagrams.

This Paper is organized as follows. In Sects.~\ref{s:MF}, \ref{insta}, \ref{sta} I review the results \cite{AM15,AM17a}
which form a background for further investigations. Sect.~\ref{s:gamma} is devoted to the computation of
the mean-field value of the string susceptibility index $\gamma_{\rm str}=1/2$ and its comparison to the
perturbative value $\gamma_{\rm str}=1$.
In Sect.~\ref{s:fluctua} I formulate a general procedure for expanding
about the mean field and describe Pauli-Villars' regulation for computing the energy-momentum tensor and its
trace anomaly, which does not rely on approximating the involved determinants
by (the exponential of) the conformal anomaly.
It Sect.~\ref{S:central} I compute the total central charge of the system
 in the mean-field approximation and show that it vanishes for
$2<d<26$.
Sect.~\ref{s:semicla} is devoted to the ``semiclassical'' expansion about the mean field.
I show that logarithmic infrared divergences which might spoil conformal invariance are mutually canceled.
The results obtained and tasks for the future are discussed in Sect.~\ref{s:discu}.
Some explicit computations are presented in Appendices~\ref{appA}, \ref{appB} by using the Mathematica programs from Appendix~\ref{appM}.

\section{The mean-field ground state\label{s:MF}}

We consider a closed bosonic string in target space with one  compactified dimension
of circumference  $\beta$. The string wraps once around
this compact dimension and propagates through the distance $L$.
The string world-sheet has thus topology of a cylinder.
There is no tachyon for such a string configuration,
if $\beta$ is larger than a certain value to guarantee that  the classical energy of
the string dominates over the energy of zero-point fluctuations.

The Nambu-Goto string action is given by the area of the 
surface embedded in target space.
It is highly nonlinear in the embedding-space coordinate $X^\mu$.
To make it quadratic in $X^\mu$, 
we rewrite it, introducing a Lagrange multiplier $\lambda^{ab}$
and an independent intrinsic metric $\rho_{ab}$, as%
\footnote{We denote $\det  \rho=\det  \rho_{ab}$ and $\det  \lambda=\det  \lambda^{ab}$.}
\be
S
=K_0 \!\int \d^2\omega\,\sqrt{\det \partial_a X \cdot \partial_bX}=
K_0 \int \d^2\omega\,\sqrt{\det  \rho} 
+\frac{K_0}2 \int \d^2\omega\, \lambda^{ab} \left( \partial_a X \cdot \partial_bX -\rho_{ab}
\right), 
\label{aux}
\ee
where $K_0$ stands for the bare string tension.
The equivalence of the two formulations can be proven by  path integrating over
the functions $\lambda^{ab}(\omega)$ and $\rho_{ab}(\omega)$ which
take on imaginary and real values, respectively.

It is convenient to  choose the world-sheet coordinates  $\omega_1$ and $\omega_2$
 inside an $\omega_L\times \omega_\beta$ rectangle in the parameter space. Then
the classical solution $X^\mu_{\rm cl}$ minimizing the action \rf{aux} 
 linearly depends on $\omega$
\be
X^1_{\rm cl}=\frac {L}{\omega_L} \omega_1,\qquad X^2_{\rm cl}=\frac {\beta}{\omega_\beta} \omega_2.
\label{Xcla}
\ee 
The classical value of $\rho_{ab}$ coincides with the classical induced metrics 
\be
\left[\rho_{\rm cl}\right] _{ab}=
\partial_a X_{\rm cl} \cdot \partial_b X_{\rm cl} = {\rm diag}\left( \frac{L^2}{\omega_L^2},
\frac{\beta^2}{\omega_\beta^2}\right) 
\label{rhocla}
\ee
which becomes diagonal for
\be
\omega_\beta =\frac \beta L \omega_L.
\label{diacla}
\ee
The classical value of $\lambda^{ab}$ reads
\be
\lambda^{ab}_{\rm cl}=\rho^{ab}_{\rm cl} \sqrt{\det \rho_{\rm cl}}
\label{lacla}
\ee
and simplifies to $\lambda^{ab}_{\rm cl}=\delta^{ab}$ if \eq{diacla} is satisfied. 

We apply the path-integral  quantization to
account for quantum fluctuations of the $X$-fields by splitting
$X^\mu=X^\mu_{\rm cl} +X^\mu_{\rm q}$ and then  performing  the Gaussian path
integral over $X^\mu_{\rm q}$. 
We thus obtain the action,
governing the fields $\lambda^{ab}$ and $\rho_{ab}$,
\bea
S_{\rm ind}&=& K_0 \int \d^2\omega\,\sqrt{\det \rho} 
+\frac{K_0}2 \int \d^2\omega\, \lambda^{ab} \left( \partial_a X_{\rm cl } 
\cdot \partial_bX_{\rm cl } 
 -\rho_{ab} \right) 
+ \frac{d}{2}  \tr \log (-{\cal O}),  \non 
{\cal O}&  = &\frac1 {\sqrt{\det \rho} }  \partial _a \lambda^{ab} \partial_b.
\label{aux1}
\eea
The operator ${\cal O}$
 reproduces the usual two-dimensional  Laplacian $\Delta$ for  $\lambda^{ab}$
given by the right-hand side of \eq{lacla}.
Its determinant is to be computed with the Dirichlet boundary condition imposed.
Quantum observables are determined by the path integral over
$\lambda^{ab}$ and $\rho_{ab}$ with the action \rf{SpS}, which
runs as is already mentioned over imaginary  $\lambda^{ab}(\omega)$ and real $\rho_{ab}(\omega)$.
The action \rf{aux1} is often called the {\em induced}\/ (or {\em emergent}) action to
be distinguished from the {\em effective}\/ action which is usually associated 
with slowly varying fields in the low-momentum limit.

It is convenient  to fix the conformal gauge when
$\rho_{ab}=\rho \delta_{ab}$, so that $\sqrt{\det{\rho}}=\rho$.
Then the log of the determinant of the ghost operator~\cite{Pol81}
\be
\big[{\cal O}_{\rm gh}\big]^a_b= \Delta^a_b - \frac 1{2} (\Delta^a_b \log \rho)
\label{Ogh}
\ee
is to be added to the induced action \rf{aux1} [or \rf{auxP} below].
The operator \rf{Ogh} acts on two-dimensional vector functions 
whose one component obeys the Dirichlet boundary condition and
the other obeys the  Robin boundary condition \cite{DOP82,Alv83}. 
The subtleties associated with the boundary conditions will be inessential
both for the matter and ghost determinants for $L\gg \beta$ when only the bulk terms survive.

We shall describe in Sect.~\ref{s:fluctua} how to accurately compute the determinants using
the Pauli-Villars regularization but let us assume for a moment that 
$\lambda^{ab}(\omega)=\blambda \delta^{ab}$ and
$\rho_{ab} (\omega)=\brho \delta_{ab}$ with constant 
$\blambda$ and $\brho$. As we shall  see these constant values 
of $\blambda$ and $\brho$ are what is needed
for the mean-field approximation.

The computation of the matter and ghost determinants 
for constant $\blambda$ and $\brho$ is presently
an 
exercise in string theory courses with the result
\be
S_{\rm eff}
=\frac {K_0}2 \blambda \left( \frac {L^2}{\omega_L^2}+ \frac {\beta^2}{\omega_\beta^2}
\right)\omega_L\omega_\beta
 +K_0\big(1 -\blambda\big)\brho\,\omega_L\omega_\beta -\left(\frac{d}{2\blambda}-1 \right)
 \Lambda^2\brho\,\omega_L\omega_\beta -\frac{\pi(d-2)}{6} \frac{\omega_L}{\omega_\beta} 
\label{SpS}
\ee
 for $L\gg \beta$.
Here $\Lambda^2$ cuts off eigenvalues of the operators involved.
The first and second terms on the right-hand side are classical contributions, while
the sign of the third term is negative for $d/\blambda>2$ to comply with positive entropy.
Technically, it comes as the product of the eigenvalues divided by $\Lambda$, where every multiplier is
less than 1. The last term is known as the L\"uscher term which is due to the Casimir energy
of zero-point fluctuations. Its negative sign is intimately linked to the presence of the tachyon.

The next step is to minimize \rf{SpS} over $\blambda$, $\brho$ to find the mean-field configuration
which describes the string  ground state. The difference from the classical ground state \rf{Xcla},
\rf{rhocla}, \rf{lacla} is that we now minimize the  action, taking into account the determinants
coming from $X^\mu$ and ghosts, while the classical (perturbative) ground state minimizes the 
classical action. Additionally, similarly to the classical case we have to minimize  \rf{SpS} over
the ratio $\omega_\beta/\omega_L$ which plays the role of the modular parameter of the cylinder.
This guarantees that $\rho_{ab}$ and $\lambda^{ab}$ are diagonal as is required by the 
conformal gauge. We shall return to this issue soon.

The minimum of \rf{SpS} is remarkable simple \cite{AM15,AM17a}
\begin{subequations}
\bea
\blambda&=&\frac{1}2 +\frac{\Lambda^2}{2K_0} +
\sqrt{\frac 14\left(1 +\frac{\Lambda^2}{K_0}\right)^2 -\frac{d\Lambda^2}{2K_0}},~~
\label{newC} \\
\brho&=& 
\frac{\left(\beta^2-\frac{\pi(d-2)}{6K_0\C}\right)}{\omega_\beta^2}
\frac \C {\sqrt{ \left(1 +\frac{\Lambda^2}{K_0}\right)^2 -\frac{2d\Lambda^2}{K_0}}},
\label{newrho} \\
\omega_\beta&=& \frac{\omega_L}{L}\sqrt{\beta^2-\frac{\pi(d-2)}{3K_0\C}}.
\label{newbeta}
\eea
\label{mmff}
\end{subequations}
The  value of the  action \rf{SpS} at the minimum \rf{mmff} is
\be
S_{\rm mf}= K_0 \C L\sqrt{\beta^2-\frac{\pi(d-2)}{3K_0\C}}.
\label{Smf}
\ee

The meaning of the above minimization procedure is clear: we have constructed a saddle-point
approximation to the path integral, which takes into account an infinite set of diagrams of 
perturbation theory 
about the classical vacuum. 
This approach is quite similar to that%
\footnote{See e.g.\ the book~\cite{Mak02}.}
 in the two-dimensional $O(N)$ sigma-model, 
where one sums up bubble diagrams of the $1/N$-expansion 
by introducing the Lagrange multiplier $u$ to resolve the constraint $\vec n^2=1$.
After integration over the fields $\vec n$ one obtains an induced action as 
a functional of $u$, whose
minimum determines  the exact vacuum state as  $N\to\infty$. 
For  finite $N$ the  fluctuations of $u$ about this mean-field  vacuum have 
to be included, but they are small even at $N=3$ because, roughly speaking, 
there is only one $u$ while the induced action is of order $N$, \ie
large as is needed for a saddle point. Alternatively, the perturbative vacuum 
$\vec n_{\rm cl}=(1,0,\ldots,0)$  possesses an $O(N-1)$ symmetry 
rather than the $O(N)$ symmetry as the saddle-point vacuum does
and  the fields $\vec n$  fluctuate strongly. 
For our case the number of fields $X^\mu$ in the sigma model \rf{aux} is $d$, so the saddle-point 
is justified by $K_0\sim d\to\infty$.
At finite $d$ the saddle-point solution \rf{mmff} is associated with the mean-field approximation.

The minimization of the  action \rf{SpS} over $\omega_\beta/\omega_L$ can be now
understood as follows. In the mean-field approximation we consider 
the  action to be large,
doing all integrals by the saddle point, including the integral over the modular parameter, which is
present for the cylinder topology.

A few comments concerning the solution \rf{mmff} are in order:
\begin{itemize}
\item
Equation \rf{newC} is well-defined if the bare string tension $K_0>K_*$ given by
\be
K_*=\left(d-1+\sqrt{d^2-2d} \right)\Lambda^2.
\label{K*}
\ee
At this critical value of  $K_0$ the square root in \rf{newC} vanishes.

\item
The classical vacuum \rf{Xcla},
\rf{rhocla}, \rf{lacla} is recovered by \rf{mmff} as $K_0\to\infty$, while the expansion in $1/K_0$ 
makes sense of the semiclassical (perturbative) expansion about this vacuum.
The usual one-loop results are recovered to order $1/K_0$.

\item
The large-$d$ ground-state energy~\cite{Alv81},%
\footnote{The original computation \cite{Alv81} used the Nambu-Goto string.
How the same result can be obtained for the Polyakov string is shown 
in \cite{Mak11}.}
where an analytic regularization was used,
are recovered by \eq{Smf} for $\Lambda^2=0$. Analogously, the ground-state energy obtained 
by the old canonical quantization \cite{Arv83} is reproduced by our mean-field approximation.
This is not surprising because fluctuations of $\rho$ are ignored in the  canonical quantization.

\item
Equation~\rf{Smf} is well-defined for $\beta$ larger than $\sqrt{\pi(d-2)/3K_0\C}\sim 1/\Lambda$,
but becomes imaginary otherwise.
The singularity was linked \cite{Alv81,Arv83,Ole85} to the tachyon mass squared.

\item
The metric \rf{newrho} becomes infinite when $K_0\to K_*$ given by \eq{K*}.
This is crucial for constructing the scaling limit.

\end{itemize}

At the classical level $\rho_{\rm cl}$ coincides with the induced metric as is displayed in \rf{rhocla}.
In the mean-field approximation it is superseded by
\be
\brho_{ab}=\LA \partial_a X \cdot \partial_b X \RA,
\label{ws}
\ee
where the average is understood in the sense of the path integral over $X^\mu$. 
Equation~\rf{ws} follows from the minimization of the effective action over $\lambda^{ab}$.
Thus, in the mean-field approximation $\brho$ coincides with the averaged induced metric.


\section{Instability of the classical vacuum\label{insta}}

The usual semiclassical (or one-loop) correction to the classical ground-state energy
due to zero-point fluctuations~\cite{BN73}
is described in textbooks. The sum of the two reads
\be
S_{1l}=\left[K_0-\frac{(d-2)}2 \Lambda^2 \right] L\beta -\frac{\pi(d-2)}{6} \frac{L}{\beta}  .
\label{S1l}
\ee
To make the bulk part of \rf{S1l} finite, it is usually introduced the renormalized string tension
\be
K_R=K_0-\frac{(d-2)}2 \Lambda^2 
\ee
which is kept finite as $\Lambda\to\infty$. Then it is assumed that it works order by
order of the perturbative expansion about the classical vacuum, so that $K_R$ can be
made finite by fine tuning $K_0$.

We see however from \eq{Smf} how it may not be case. The right-hand side of \eq{Smf}
never vanishes with changing $K_0$. The point of view on \eq{S1l} should be that
for $d>2$  the one-loop
correction simply lowers the energy of the classical ground state which 
therefore may be unstable.

As we show in the next Section, the action~\rf{SpS} indeed increases if we add a constant imaginary
addition $\delta \lambda$ to $\blambda$. However,  the sum of the two linear in $\brho$ terms in
\eq{SpS} vanishes for $\blambda$ given by \eq{newC}, so the action does {\em not}\/ depend on 
$\brho$ at the minimum. This reminds a valley in the problem of spontaneous symmetry breaking.

To investigate it, 
we proceed in the standard way, adding to the action the source term
\be
S_{\rm src}=\frac {K_0}2 \int \d^2 \omega \, j^{ab} \rho_{ab}
\ee
and defining the field
\be
\rho_{ab}(j)= -\frac {2}{K_0} \frac \delta {\delta j^{ab}}\log Z .
\ee
Minimizing the action with the source term added
for constant $j^{ab}=j \delta ^{ab}$, we find \cite{AM17a}
\be
\C(j)=\frac{1}2 \left(1+ j +\frac{\Lambda^2}{K_0}\right)+
\sqrt{\frac 14\left(1 +j+\frac{\Lambda^2}{K_0}\right)^2 -\frac{d\Lambda^2}{2K_0}}
\label{newCj} 
\ee
and
\be
\brho(j) 
=\frac12+
\frac { 1+j+\frac{\Lambda^2}{K_0} }{\sqrt{\left(1 +j+\frac{\Lambda^2}{K_0}\right)^2 -\frac{2d\Lambda^2}{K_0}}}
\label{rhoj}
\ee
in the mean-field approximation
for $\omega_L=L$ and $\omega_\beta=\beta \gg 1\sqrt{K_0}$. 
Inverting \eq{rhoj}, we obtain
\be
j(\brho)=-1-\frac{\Lambda^2}{K_0}+\sqrt{\frac{d\Lambda^2}{2K_0}}
\frac{(2\brho-1)}{ \sqrt{\brho(\brho-1)}}.
\label{jjrr}
\ee

To understand the properties of the vacuum,
we compute an ``effective potential'' by performing the Legendre transformation
\be
\Gamma(\brho)=-\frac{1}{K_0L\beta} \log Z
- j (\brho) \brho ,
\label{Ga}
\ee
like in the studies of 
symmetry breaking in quantum field theory. 

In the mean-field approximation  we then obtain
\be
\Gamma(\brho)=
\left( 1+\frac{\Lambda^2}{K_0} \right) \brho -\sqrt{\frac{2d \Lambda^2}{K_0}  \brho(\brho-1)}.
\label{bGa}
\ee
Note that
\be
-\frac{\partial \Gamma(\brho)}{\partial \brho}=j(\brho)
\label{10}
\ee
with $j(\brho)$ given by \eq{jjrr} as it should.

Near the classical  vacuum we have $0<\brho-1\ll 1$ and the potential
\rf{bGa} decreases with increasing
$\brho$ because the second term on the right-hand side has
 the negative sign, demonstrating an 
instability of the classical vacuum. If 
$K_0>K_* $ given by \eq{K*},
the potential \rf{bGa} linearly increases with $\brho$ for large $\brho$ and thus has a (stable) minimum at 
\be
\brho(0) =\frac 12 + \frac{1+\frac{\Lambda^2}{K_0}}
{2\sqrt{\left(1+\frac{\Lambda^2}{K_0}\right)^2-\frac {2d\Lambda^2}{K_0}}}
\label{bbrr}
\ee
which is the same as \rf{newrho} for $\omega_\beta=\beta\gg1/\sqrt{K_0}$.
Near the minimum we have
\bea
\Gamma(\brho)&=&\left[ \left(1+\frac{\Lambda^2}{K_0}\right)^2-\frac {2d\Lambda^2}{K_0} \right]^{1/2}+  \frac{K_0}{2d \Lambda^2} 
\left[ \left(1+\frac{\Lambda^2}{K_0}\right)^2-\frac {2d\Lambda^2}{K_0} \right]^{3/2}\!
(\Delta \brho)^2  +
{\cal O}\left((\Delta \brho)^3\right), \non
\Delta \brho&=&\brho-\brho(0).
\eea
The coefficient in front of the quadratic term is positive for $K_0>K_*$ which explicitly
demonstrates the (global) stability of the mean-field minimum \rf{newrho}.

The situation is different for $d<2$, where quantum corrections increase the vacuum energy.
For this reason the classical vacuum is energetically favorable to the mean-field one.
It is explicitly seen for $d<0$ from \eq{bGa} where $\brho-1$ has to be negative. 
The function $\Gamma(\brho)$ then increases with decreasing  $\brho$ near $\brho=1$ and the 
mean-field solution is a maximum, not a minimum.

The conclusion of this Section is that the classical vacuum is not stable for $d>2$ where the
mean-field vacuum is energetically favorable. This reminds spontaneous generation of $\brho$ in
quantum field theory.
 The situation is opposite for $d<2$, where the classical vacuum has lower energy than the
mean-field vacuum.

\section{Stability of the mean-field vacuum\label{sta}}

Let us now consider stability of the mean-field vacuum under wavy fluctuations, when
\be
\rho(\omega)=\brho+\delta \rho(\omega), \qquad 
\lambda^{ab}(\omega)=\blambda \delta^{ab}+\delta \lambda^{ab}(\omega)
\ee
with $\omega$-dependent $\delta\rho$ and $\delta\lambda$.

The divergent part of the effective action reads~\cite{AM17a}
\bea
S_{\rm div}&=&\int\d^2 \omega
\left[ \frac{K_0}2 \lambda^{ab} \partial_a X_{\rm cl}\cdot \partial_b X_{\rm cl}+
K_0  \rho\left(1 -\frac12  \lambda^{aa} \right) \right.  \left .-
\frac {d \Lambda^2 \rho}{2\sqrt{\det{\lambda}}}
+ \Lambda^2   \rho \right],  \non
\lambda^{aa}&=&\lambda^{11}+\lambda^{22}.
\label{cla}
\eea
For constant $\lambda^{ab}=\blambda \delta^{ab}$
and $\rho=\brho$ this reproduces the divergent part of \eq{SpS} above.

Expanding to quadratic order in fluctuations
\bea
&&\sqrt{ \det(\blambda \delta^{ab}+\delta \lambda^{ab})}=\blambda+\frac12 \delta \lambda^{aa}
-\delta \lambda_2 
+{\cal O}\left((\delta \lambda)^3\right), \non
&&\delta \lambda_2 =\frac1{8\blambda} (\delta \lambda_{11}-\delta \lambda_{22})^2+ 
\frac 1{2 \blambda}(\delta \lambda_{12})^2,
\label{3001}
\eea
we find from \rf{cla} 
\be
S^{(2)}_{\rm div} = -\frac{d\Lambda^2 \brho}{2\C} \int \d^2 \omega\,\delta \lambda_2
-\left(K_0-\frac{d\Lambda^2}{2\C^2}\right)\!\int \d^2 \omega \, \delta \rho \frac{ \delta \lambda^{aa}}2
-\frac{d\Lambda^2 \brho}{2\C^3}\int \d^2 \omega \left( \frac{ \delta \lambda^{aa}}2\right)^2 .
\label{Sdi}
\ee

The first term on the right-hand side of \eq{Sdi} plays a very important role for dynamics
of quadratic fluctuations. Because the path integral over $\lambda^{ab}$ goes
 parallel to imaginary axis,
\ie  $\delta \lambda^{ab}$ is pure imaginary, 
 the first term is always {\em positive}. 
Moreover, 
its exponential plays the role of a (functional) delta-function as $\Lambda\to\infty$,
forcing $\delta \lambda^{ab}=\delta \lambda \,\delta^{ab}$.
The same is true for a constant part of $\delta \lambda^{ab}$.

For the effective action 
to the second order  in fluctuations we then find the following quadratic form:
\be
\delta S_2=\int \frac{\d^2p}{(2\pi)^2} \left[
A_{\rho\rho} \frac{\delta \rho(p)\delta \rho(-p) }{\bar \rho^2} +
2A_{\rho\lambda} \frac{\delta \rho(p) \delta \lambda(-p)}{\bar  \rho \blambda} 
+A_{\lambda\lambda} \frac{\delta \lambda(p)\delta \lambda(-p) }
{\blambda^2} \right],
\label{191}
\ee
where
\begin{subequations}
\bea
A_{\rho\rho} &= &
\frac {(26-d) p^2}{96\pi},\\
A_{\rho\lambda}&=& -\frac12\left(K_0-\frac{d\Lambda^2 }{2\C^2}\right)\brho \C
-\frac {d p^2}{48\pi}, \\
A_{\lambda\lambda}&=& -\frac{d \Lambda^2\bar \rho} {2\C} -\frac {d p^2}{32\pi}
\log \frac{cp^2}{\Lambda^2 \brho}.
\label{AAAA}
\eea
\label{MMMM}
\end{subequations}
Here $c$ is a regularization-dependent constant.

In the scaling limit, where \cite{AM15,AM17a} 
\be
K_0\to K_*+ \frac{K_R^2}{2\Lambda^2 \sqrt{d^2-2d}}
\label{scal}
\ee
as $\Lambda\to\infty$ keeping   the renormalized string tension  $K_R$ fixed,
we have
\be
K_0-\frac{d \Lambda^2}{2\C^2}\to K_R \left(1+\sqrt{1-\frac2d} \right),
\label{scal1}
\ee
so only $A_{\lambda\lambda}$ diverges as $\Lambda^2$. Therefore, typical $\delta\lambda\sim1/\Lambda$
so that $\lambda^{ab}$ is localized at the value
\be
\blambda^{ab} =\C\delta^{ab} .
\label{cbla}
\ee
This is quite similar to what is shown in the book~\cite{Pol87} for the fluctuations
about the classical vacuum.
Thus only $\rho$ fluctuates.

Equation \rf{cbla} holds in the conformal gauge, where  $\rho^{ab} \sqrt{\det \rho} =\delta^{ab} $.
In the general case
the field $\lambda^{ab}(\omega)$ is localized at the value
\be
\blambda^{ab}=\C \rho^{ab} \sqrt{\det \rho} ,
\label{bla}
\ee
where $\C$ is constant for the world-sheet parametrization in use.

We can therefore rewrite the right-hand side of \eq{aux} in the scaling limit as
\be
S=K_0 (1-\C)\int \d^2\omega\,\sqrt{\det  \rho} 
+\frac{K_0 \C}2 \!\int \d^2\omega\,  \sqrt{\det \rho}\,
\rho^{ab}\partial_a X \cdot \partial_b X ,
\label{auxP}
\ee
which reproduces the Polyakov string formulation~\cite{Pol81} for $\C=1$.
As shown in \cite{AM15} the action \rf{auxP} is consistent only for
a certain value of $\C$ which is regularization-dependent.
One has $\C=1$ for the zeta-function regularization
but  $\C<1$ for the proper-time regularization or the Pauli-Villars  regularization.

A subtlety with the computation of the determinants in the conformal gauge
is that $X^\mu$ and $\rho$ do not interact in the action~\rf{auxP}
since 
\be
S=K_0 (1-\C)\int \d^2\omega\,  \sqrt{\hat g}\,\rho 
+\frac{K_0 \C}2 \!\int \d^2\omega\,   \sqrt{\hat g}
{\hat g}^{ab}\,\partial_a X \cdot \partial_b X 
\label{auxPc}
\ee
in the conformal gauge $\rho_{ab}=\hat g_{ab}\rho$. Here $ {\hat g_{ab}}$ is a fiducial metric which 
we can set $ {\hat g_{ab}}=\delta_{ab}$ without loss of generality.

But the dependence of the determinants on $\rho$ appears because the world-sheet
regularization 
\be
\varepsilon=\frac 1{\Lambda^2 \sqrt{ \det \rho} }=
\frac 1{\Lambda^2  \rho }
\ee
depends on $\rho$ owing to diffeomorphism invariance.
For smooth $\rho$ the determinants are given by the usual conformal anomaly~\cite{Pol81}.
An advantage of using the Pauli-Villars regularization in the conformal gauge is that 
the implicit dependence on the metric becomes explicit as is described in Sect.~\ref{s:fluctua}.

Integrating over the matter and ghost fields, we arrive for $\hat g_{ab}=\delta_{ab}$ 
to the induced action 
\bea
S_{\rm ind}&=&
\frac{K_0 \C}2 \!\int \d^2\omega\,  
\delta^{ab}\partial_a X_{\rm cl}  \cdot \partial_b X_{\rm cl} +K_0 (1-\C)\int \d^2\omega\, \rho \non && +
 \frac d2 \tr \log \left( -\frac {\C}\rho \partial^2
\right)\Big|_{\rm  reg} - \frac 12   \tr \log \Big(\big[{-\cal O}_{\rm gh}\big]^a_b\Big)\Big|_{\rm reg} ,
\label{effgen}
\eea
where the ghost operator is displayed in \eq{Ogh}.
Evaluating the determinants, we find for smooth $\rho$
\bea
S_{\rm eff}&=&\frac{K_0 \C}2 \left(\frac{L^2 \omega_\beta}{\omega_L} +\frac{\beta^2 \omega_L}{\omega_\beta}
\right)+\left[K_0 (1-\C)-\left(\frac{d}{2\C}-1\right) \Lambda^2 \right]\int \d^2\omega\, \rho  \non
&&+\frac {26-d}{96 \pi} \int \d^2\omega\,  (\partial _a \log \rho )^2
\label{effPc}
\eea
which for $\C=1$ reproduces the usual result.

We see from \eq{effPc} (as well as from \eq{191} with $\delta \lambda=0$) that the action,
describing fluctuations of the metric, is positive
only for $d<26$ and becomes negative if $d>26$. Thus,
as far as the local stability of the  action under wavy fluctuations is concern, it is the same about
the mean-field vacuum as about the
usual classical vacuum.
This instability is probably linked to the presence of negative-norm states for $d>26$
\cite{Bro72,GT72}.

\section{The string susceptibility index\label{s:gamma}}

A very important characteristics of the string dynamics is
the string susceptibility index $\gamma_{\rm str}$ which 
characterizes the string entropy and  is determined from the preexponential 
in the number of surfaces of fixed area $A$ by
\be
\e^{-F(A)}\equiv \LA \delta\Big(  \int \d^2 z\,\rho - A \Big) \RA \stackrel{A\to\infty}\propto A^{\gamma_{\rm str}-2} \e^{C A} ,
\label{ggstr}
\ee
where $C$ is a nonuniversal constant. $F(A)$ on 
the left-hand side has the meaning of the Helmholtz free energy
of a canonical ensemble at fixed area $A$.
Introducing the Lagrange multiplier, we rewrite \rf{ggstr} as
\be
\LA \delta\Big(  \int  \d^2 z\,\rho - A \Big) \RA = \LA
 \int \d j \,\e^{j \left(\int  \d^2 z\,\rho-A\right)}\RA,
\label{FFF}
\ee
where the integral over $j$  runs parallel to the imaginary axis.
This $j$ is the same as introduced in Sect.~\ref{insta} except for the integral over $j$.

Let us first consider the integrand. The saddle-point solution is given by \eq{rhoj}.
Then the integrand in \rf{FFF} has an extremum at $j(A)$ given 
by \eq{jjrr} with $\brho$ substituted by $A/A_{\rm min}$, ${A_{\rm min}}=L\beta$.
Expanding about the extremum, we find~\cite{AM17c}
\bea
\frac {jA} {K_0A_{\rm min}}- \C(A) &= &\sqrt{\frac{2d \Lambda^2}{K_0}}
 \sqrt{ \frac{ A}{A_{\rm min}} \left(\frac A{A_{\rm min}}-1\right) }
-\frac A{A_{\rm min}}\left( 1+\frac{\Lambda^2}{K_0} \right) \non &&+\sqrt{\frac{2K_0}{d\Lambda^2}}
\left[\frac{ A}{A_{\rm min}} \left(\frac A{A_{\rm min}}-1\right) \right]^{3/2} \left(\Delta j\right)^2.
\label{15}
\eea
The integral over $\Delta j=j-j(A)$ goes along the imaginary axis
and thus converges. 
For $F(A)$ we obtain
\be
F(A)= \sqrt{{2d \Lambda^2}{K_0}} \sqrt{ A (A-{A_{\rm min}}) }-A (K_0+\Lambda^2)+
\frac3{4 } \log \left[ A (A-{A_{\rm min}}) \right] + {\rm const.}
\label{16}
\ee

According to the definition \rf{ggstr} of the string susceptibility index, we expect
\be
 F(A)={\rm regular}+ \left(  2-\gamma_{\rm str} \right) \log \frac{A}{A_{\rm min}}
\label{regular}
\ee
for ${A}\gg A_{\rm min}$. Comparing with  \rf{16},
this determines $\gamma_{\rm str}=1/2$.
It can be shown~\cite{AM17c}  that the one-loop correction contributes only to the 
regular part of $F(A)$ and does not change the singular
part that gives $\gamma_{\rm str}=1/2$. This value can be exact  because it is linked
only to the emergence of the square-root singularity which is not changed by higher
orders.

We are to compare the mean-field  result for  $\gamma_{\rm str}$
 with the one-loop computation of \rf{ggstr} about the classical vacuum which is
almost trivially done by changing $\rho\to A\rho$ and gives
\be
\rf{ggstr}\propto A^{-1} \e^{(d-2)\Lambda^2A /2}
\ee
resulting in $\gamma_{\rm str}=1$. We can compare it with the formula of the $d\to-\infty$ expansion \cite{Zam82}
generalized to an arbitrary genus in \cite{CKT86,KK86}. Since we deal with the worldsheet having topology of
a cylinder which has two boundaries, its Euler character equals 0 like for a torus. This explains
why there is no $d$-dependence of $\gamma_{\rm str}$. We have got
$\gamma_{\rm str}=1$ rather than $\gamma_{\rm str}=2$
as in \cite{CKT86,KK86} because we deal with an open rather than a closed string.

The discrepancy between the obtained mean-field value  $\gamma_{\rm str}=1/2$ and the perturbative value
 $\gamma_{\rm str}=1$ is due to the fact that the vacua are different. The former applies for $2<d<26$, while the
latter applies for $d<2$.

\section{Fluctuations about the mean-field\label{s:fluctua}}

The instability of the effective action 
for $d>26$ implies that we cannot straightforwardly make a systematic $1/d$ expansion as
$d\to +\infty$. This is in contrast to the $d\to -\infty$ limit which comes along with the usual perturbative
expansion because the vacuum is then just classical.
The usual semiclassical expansion as $d\to -\infty$ cannot be extended to $d>2$ because
the vacuum states are different for $d<2$ and $d>2$.

To go beyond the mean field for $2<d<26$, we define the partition function
\be
Z[h]=\int {\cal D}\rho \e^{-S_{\rm ind}/h}
\label{Zh}
\ee
with $S_{\rm ind}$ given by \eq{effgen}.
Here we have introduced an additional parameter $h$ to control the ``semiclassical'' expansion
about the mean field which plays the role of a ``classical'' vacuum. 
This procedure makes sense of the change $d\to d/h$ for the number of the $X$-fields and
simultaneously $2\to 2/h$ for the number of the ghost fields. Diagrammatically, this mean 
field corresponds  to summing up bubbles of both matter and ghosts.
The mean-field approximation is associated with $h\to 0$, while the expansion 
about the mean field goes in $h$.
Diagrams with $l$ loops are then proportional to $h^l$.
In reality $h=1$ but we can expect that the actual expansion 
parameter is $6h/(26-d)$ like in the usual semiclassical expansion as $d\to -\infty$.
Then the expansion can make sense for $d=4$.

The action in \eq{Zh} is given by \rf{effgen}. For the Pauli-Villars regularization 
the determinants are regularized  by the ratio
of massless to massive determinants~\cite{AM17c}
\be
\det(-{\cal O})\big|_{\rm reg}\equiv\frac{\det(-{\cal O})\det(-{\cal O}+2M^2)}{\det(-{\cal O}+M^2)^2},
\label{newR}
\ee
so that
\be
 \tr\log(-{\cal O})\big|_{\rm reg}= -
\int_{0}^\infty \frac{\d \tau}{\tau} \,\tr \e^{\tau{\cal O}}
\left(1-\e^{-\tau M^2}\right)^2
\label{PV22}
\ee
is convergent. Here $M\to\infty$ is the regulator mass which is related to $\Lambda$ in the above 
equations by
\be
\Lambda^2 = \frac{M^2}{2\pi}\log 2.
\label{M2}
\ee
For the proper-time regularization  we have instead
\be
 \det(-{\cal O})\big|_{\rm reg}= -
\int_{a^2}^\infty \frac{\d \tau}{\tau} \,\tr \e^{\tau{\cal O}},\qquad \Lambda^2=\frac1{4\pi a^2}.
\label{ptR}
\ee

We have added in \rf{newR} the ratio of the
determinants for the masses $\sqrt{2}M$ and $M$ to cancel the
logarithmic divergence at small $\tau$, because the Seeley 
expansion
\be
\LA \omega \Big| \e^{\tau {\cal O} }
\Big| \omega \RA =
\frac1{4\pi \tau}    
\frac{1}{\sqrt{\det{\lambda}}} +\frac {R}{24\pi}+\ldots
\label{Seeley}
\ee
starts with the term  $1/\tau$.
This is specific to the two-dimensional case.

The massive determinants in \eq{newR} can also be represented as path integrals of the type
\be
\det \left(  -\frac \C\rho \partial^2 + M^2 \right)^{-d/2}
=\int {\cal D} X^\mu_M
\e^{-\frac{K_0}2\int \d^2 \omega\, \left(\blambda\delta^{ab} \partial_a X_M \cdot \partial_b X_M + 
M^2 \rho X_M\cdot X_M \right)}
\ee
over the fields  $X_M(\omega)$ with normal statistics or  $Y_M(\omega)$ 
with ghost statistics and the double number of components.
We can explicitly add these regulator fields to the action \rf{auxPc} to get
\bea
S&=&K_0 (1-\C)\int \d^2\omega\, \rho +\frac{K_0}2 \int \d^2\omega\,   \left[
 \C\,\partial_a X \cdot \partial_a X\right. \non && \left.+\sum_{i=1}^2 \left(\C\,
\partial_a Y_M^{(i)} \cdot \partial_a Y_M^{(i)}  +M^2 \rho (Y_M^{(i)}) ^2 \right) +
\left(\C\,
\partial_a X_{\sqrt 2 M} \cdot \partial_a X_{\sqrt 2 M} +2M^2 \rho X^2 _{\sqrt 2 M}\right)
\right]. \non &&
\label{auxPcreg}
\eea

The path integral over the regulator fields
generates the propagator 
\be
\LA X^\mu_M(k)X^\nu_M(-k) \RA= \frac {\delta^{\mu\nu}}{K_0(\blambda k^2 + M^2 \brho)}
\ee
and the triple vertex of the
$\delta \rho X^\mu_M  X^\nu_M$ interaction
\bea
\LA\delta \rho(-p) X^\mu_M (k+p) X^\nu _M(-k) \RA_{\rm truncated}&=& - K_0  M^2 \delta^{\mu\nu } .
\label{trivertex}
\eea
The latter vanishes for $M=0$  as it should owing to conformal invariance, but 
explicitly breaks it at nonzero $M$. 
Notice that path integration over all matter fields (both $X^\mu$ and the regulators) runs with
a simple nonregularized measure. 
This makes it very convenient to derive (regularized) 
Noether's currents and to calculate their anomalies.

\begin{figure}
\includegraphics[width=12cm]{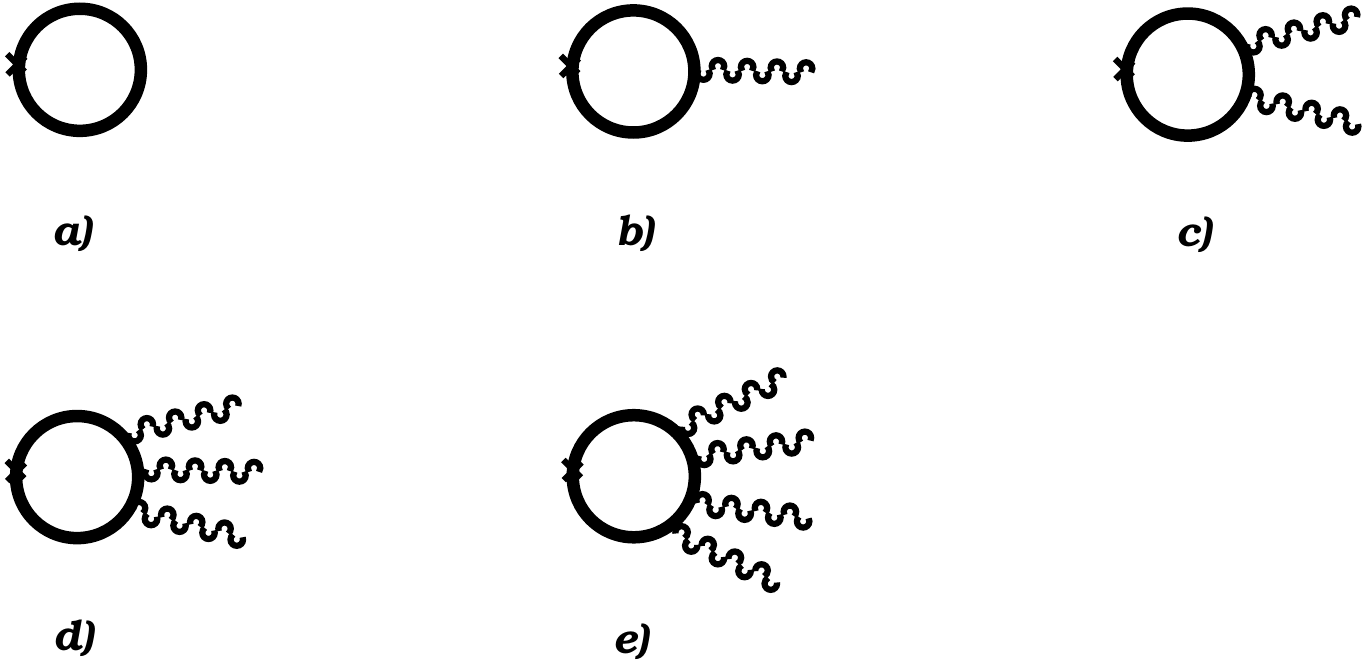} 
\caption{Diagrams contributing to the average of  \rf{Taa} or \rf{Tzzmat} over the regulator fields.}
\label{mean-F1}
\end{figure}

An instructive exercise is how to compute the usual anomaly in the trace of the energy-momentum
tensor
\be
[T_{a}^a]_{\rm mat}=4\pi K_0
\left[ 1-\C  +  \frac12 \left( \sum _{i=1,2}M^2 (Y_ M^{(i)}) ^2+ 2M^2 X _{\sqrt 2 M}^2\right)\right].
\label{Taa}
\ee
Averaging \rf{Taa} over the regulator fields, we obtain
the diagrams depicted in Fig.~\ref{mean-F1},
where the solid line corresponds to the propagator of
the regulator fields  $X_{\sqrt 2 M}$ or $Y_M$ while the wavy line
corresponds to  $\delta\rho$. We have explicitly in momentum space
\be
{\rm Fig.} ~1a=2\pi \frac{ d}{h} \int \frac{\d^2k}{(2\pi)^2}\, 
\left(\frac{2M^2}{\C k^2+2M^2\brho}-\frac{2M^2}{\C k^2+M^2\brho}
\right)=-\frac d {\C h} M^2 \log 2,
\ee
reproducing \eq{M2}, and
\bea
{\rm Fig.} ~1b&=&2\pi \frac{ d}{h} 
\int \frac{\d^2k}{(2\pi)^2}\, \left[\frac{4M^4}{(\C k^2+2M^2\brho)(\C (k-p)^2+2M^2\brho)} 
\right.\non && \left. \hspace*{.7cm}
-\frac{2M^4}{(\C k^2+M^2\brho )(\C (k-p)^2+M^2\brho)}
\right]  
\equiv\frac d {12\brho^2 h} G(p) \stackrel {p\to0}\to \frac d {12\brho^2 h}p^2,~
\label{51}
\eea
where
\be
{G(p)}=12 \left( \frac{4m^4 \,{\rm arctanh}\, \frac{p\sqrt{p^2+8m^2}}{p^2+4m^2}}{p(p^2+8m^2)} -
\frac{2m^4 \,{\rm arctanh}\, \frac{p\sqrt{p^2+4m^2}}{p^2+2m^2}}{p(p^2+4m^2)}\right) \stackrel{p\ll m}= p^2
\label{G(p)}
\ee
and for brevity we denoted
\be
m^2=M^2 \frac \brho \blambda.
\label{defm}
\ee

The effect of the diagram in Fig.~\ref{mean-F1}$c$ and the next orders 
is to complete the result to scalar curvature $R$ as is discussed in Appendix~\ref{appA}.
Adding all diagrams and using \eq{M2}, we obtain for the contribution from matter
\be
\LA[T_{a}^a]_{\rm mat}\RA=4\pi \left(K_0(1-\C)-
\frac {d}{2\C} \Lambda^2 \right) +\frac{d}{12} R .
\label{6.16}
\ee

It still remains to compute the contribution of the ghost determinant which
we also regularize by the Pauli-Villars regularization
\be
 \det \Big(\big[{-\cal O}_{\rm gh}\big]^a_b\Big)\Big|_{\rm reg}
= \frac{ \det \Big(\big[{-\cal O}_{\rm gh}\big]^a_b\Big) 
 \det \Big(\big[{-\cal O}_{\rm gh}\big]^a_b+2M^2 \delta^a_b\Big) }
{ \det \Big(\big[{-\cal O}_{\rm gh}\big]^a_b+M^2 \delta^a_b\Big)^2}.
\ee
The computation of the contribution from ghosts is pretty much similar to the 
one~\cite{Pol81,DOP82,Alv83} for the perturbative vacuum 
and adding it with \rf{6.16}
we obtain for the trace of  the total energy-momentum tensor (matter plus ghosts)
\be
\LA T_{a}^a \RA \equiv
\LA \Big([T_{a}^a]_{\rm mat}+[T_{a}^a]_{\rm gh}\Big)\RA
=4\pi \left[K_0(1-\C)-\left(\frac {d}{2\C}-1 \right)\Lambda^2 \right]+\frac{d-26}{12} R .
\label{53}
\ee
which is the same as $ \delta/\delta \rho$ acting on \rf{effPc}.
The average in this formula is over the matter and ghost fields but not over $\rho$ 
which plays the role of an external field.

For $\C$ given by \eq{newC} the divergent term vanishes, so we reproduce the usual 
conformal anomaly. The reason is that we have essentially made a one-loop calculation 
for the Polyakov-like action \rf{auxPc}
with a constant fiducial metric $ \hat \rho_{ab}=\brho \delta_{ab}$ and the result coincides with
the one about the classical vacuum because of the background independence.

\section{Computation of the central charge\label{S:central}}

If $\rho_{ab}$ is considered as a classical background metric, only matter and
ghosts contribute to the central charge of the Virasoro algebra which equals $d\!-\!26$
like in \eq{53}. Then the conformal anomaly vanishes only in $d=26$ (the critical dimension)
which reproduces the result of the old canonical quantization.
We shall now see how this is modified when quantum fluctuations of $\rho_{ab}$ are
taken into account in the mean-field approximation.

For this purpose let us compute 
the correlator  of  the two  $zz$-components of the energy-momentum
tensor
\be
T_{zz}\equiv T(z)= T_{\rm mat}(z)+ T_{\rm gh}(z).
\label{Tzz}
\ee
Classically, the $X$-field does not interact, as is already pointed out,  with the metric $\rho$
in the conformal gauge because of conformal invariance. Like in the previous Section
we shall make use of the Pauli-Villars regularization, where $T_{\rm mat}(z)$ explicitly
depends on the regularizing fields as 
\be
T_{\rm mat}(z)=2\pi K_0 \C\left(\partial_z X \cdot \partial_ z X +
\partial_z X_{\sqrt 2 M} \cdot \partial_ z X _{\sqrt 2 M}+\sum_{i=1}^2
\partial_z Y_M^{(i)} \cdot \partial_ z Y _ M^{(i)} \right).
\label{Tzzmat}
\ee

\begin{figure}
\includegraphics[width=8.4cm]{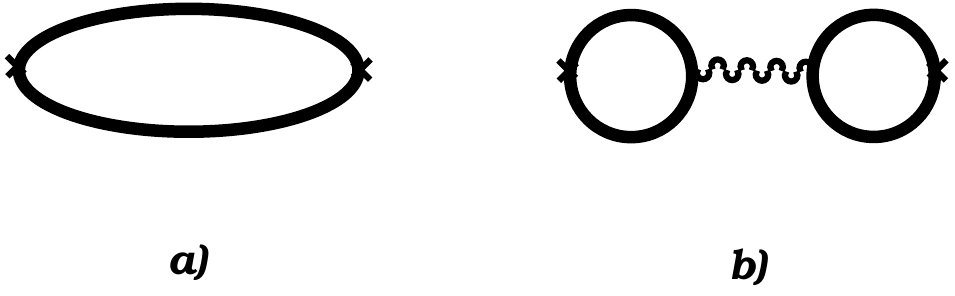} 
\caption{Diagrams contributing to the correlator $\LA T(z) T(0) \RA$  in the mean-field approximation.}
\label{mean-F2}
\end{figure}

The diagrams contributing to the correlator
$ 
\LA T(z) T(0) \RA
$ 
in the mean-field approximation are depicted in
 Fig.~\ref{mean-F2},
where the solid line corresponds to the propagator of
the field $X$ (and its regulators $X_{\sqrt 2 M}$ 
and $Y_M$) or the ghosts (and their regulators), while the wavy line
corresponds to the propagator of $\delta\rho$
\be
\LA \delta \rho(-k) \delta \rho(k) \RA =\frac {48\pi h}{(26-d) k^2}, \quad
\LA \delta \rho(z) \delta \rho(0) \RA =-\frac {12 h}{(26-d) } \log(z\bar z).
\label{rhorho}
\ee
To each closed line there is associated a factor of $({d-26})/h $
coming from summation over the matter and ghosts like in \eq{53}.

The diagram in Fig.~\ref{mean-F2}$a$ (which have a combinatorial factor of 2)  gives the usual result
\be
\LA T(z) T(0) \RA_{a)}=\frac {d-26}{2 h z^4}
\label{1a}
\ee
associated with the central charges of free fields: $d$ for matter and 26 for ghosts,
whose difference vanishes only in the critical dimension $d=26$.
Only massless fields contribute to the most singular as $z\to0$ part of the correlator 
shown in \eq{1a} via the propagator
\be
\LA X_q^\mu(z) X_q^\nu(0) \RA =-\frac 1{4\pi K_0\C} \delta^{\mu\nu}\log (z \bar z).
\ee

The diagram in Fig.~\ref{mean-F2}$b$ is usually associated with the next order 
of the perturbative  expansion about the classical vacuum
because it has two loops, but in the mean-field
approximation it has to be considered together with the diagram in  Fig.~\ref{mean-F2}$a$ 
since both are of the same order in $h$. 
We shall return soon to the discussion of this issue.
Every of the two closed loops in the diagram in  Fig.~\ref{mean-F2}$b$ involves the 
momentum-space integral
\be
2\pi \int \frac{\d^2 k}{(2\pi)^2} \, k_z(k_z-p_z) \left\{\frac{2M^2}{(k^2+2M^2)[(k-p)^2+2M^2]} -
\frac{2M^2}{(k^2+M^2)[(k-p)^2+M^2]} \right\}= \frac{p_z^2}{12},
\label{intzz}
\ee
where we have absorbed the ratio $\brho/\C$ into $M^2$ for simplicity.
The power counting predicts a quadratically divergent term like $M^2 \rho_{zz}$ in the integral~\rf{intzz},
but it vanishes in the conformal gauge. 

Each of the two closed lines is associated ether with matter (the factor of $d$) of
ghosts (the fector of $-\!26$).
Multiplying the contribution of the two loops by the propagator
\rf{rhorho}, we find for the diagram in  Fig.~\ref{mean-F2}$b$ 
\be
\LA T(z) T(0) \RA_{b)}=\frac {(d-26)}{12 h} \frac {12 h}{(26-d) }\frac {(d-26)}{12 h}
\frac 6{ z^4}=-\frac {d-26}{2 h z^4}.
\label{1b}
\ee
Notice this result is pure anomalous: it
comes entirely from the regulator fields but $M$ has canceled.
Both diagrams in Fig.~\ref{mean-F2} give a ``classical'' 
(\ie saddle-point) contribution from
the viewpoint of the mean field.
Adding \rf{1a} and \rf{1b}, we obtain zero value of the total central charge in the mean-field approximation.

The fact that the total central charge of the bosonic string is always zero in the mean-field approximation, independently on the number of the target-space dimensions
 $d$, is remarkable. Thus it always reminds the string in the critical dimension
 $d=26$.

A very similar situation occurs in the Polchinski-Strominger
approach~\cite{PS91} to the effective string theory, where the 
Alvarez-Arvis ground-state energy (same as \rf{Smf} for $\C=1$) 
was obtained from the requirement of  vanishing the central charge
at large $\beta$ to order $1/\beta$ \cite{PS91}, $1/\beta^3$~\cite{Dru04} and
$1/\beta^5$~\cite{AFK11}.
The mean-field approximation we used apparently sums up bubble graphs to 
all orders in $1/\beta$ and explicitly results in the Alvarez-Arvis formula.

\section{``Semiclassical'' correction to the mean field\label{s:semicla}}

Let us consider a ``semiclassical'' correction to the mean-field approximation which comes from averaging over
fluctuations of $\rho$ about $\brho$.

 Integrating over the matter and ghost fields (including their regulators), 
we obtain the following induced
action for the field $\delta \rho$ to quadratic order in $\delta \rho$:
\be
S_{\rm ind}^{(2)}=\frac{(26-d)}{96\pi h} \int \frac{\d^2 p}{(2\pi)^2} \delta \rho(-p) G(p) \delta \rho(p)
\ee
with $G(p)$ given by \eq{G(p)}. 

This is not the end of the story because there are diagrams with three, four, etc. $\delta \rho$'s in \rf{effgen}, whose
contributions we denote as $S_{\rm ind}^{(3)}$,  $S_{\rm ind}^{(4)}$, etc.
As is explicitly demonstrated in Appendix~\ref{appA}, it is convenient to introduce instead of $\delta \rho$ another variable
$\vp$ by
\be
 \rho(z) = \brho\e^{\varphi(z)}, \qquad
\delta \rho(z) =  \brho\left( \e^{\varphi(z)}-1 \right)
\label{drho}
\ee
and to expand in $\vp$. Then the terms higher than quadratic order in $\vp$ are mutually canceled 
in the sum 
\be
S_{\rm ind}=
S_{\rm ind}^{(2)}+S_{\rm ind}^{(3)}+\ldots 
=\frac{(26-d)}{96\pi h} \int \frac{\d^2 p}{(2\pi)^2} \vp(-p) G(p) \vp(p) +
{\cal O}(\vp^3) 
\label{Sind}
\ee
in the IR limit where all  variables $p_i$'s obey $p_i p_j\ll M^2\brho $,  
so the {\em induced}\/ action \rf{Sind} reproduces 
the usual {\em effective}\/ action for smooth $\vp (z)$. However, we consider below explicitly the case of
four $\vp$'s, where two momenta are small, $p_i^2\ll M^2 \brho$, but two other momenta are large, 
$p_i^2\sim M^2 \brho$. There is no reason to expect the cancellation in this case.

The effective action describes ``slow'' fluctuations of $\vp$ with $p^2\ll M^2\brho$ and emerges after averaging over ``fast''
 fluctuations with $p^2\sim M^2\brho$.  The quadratic part of the effective action gets then contribution from averaging higher terms in the induced  action (which are generically nonlocal), 
so we write it in the spirit of DDK  (a good review is~\cite{ZZ}) as
\be
S_{\rm eff}=\frac{1}{16\pi b^2}
\int \d^2z\, \partial_a \vp\partial_a \vp +{\cal O}(M^{-2})
\label{Seff}
\ee
with a certain constant $b^2$.
The difference between the induced action \rf{Sind} and the effective action \rf{Seff} will show up when virtual momenta 
of the propagator  $\LA \vp(-p) \vp(p) \RA $ 
in  diagrams, which emerge  after averaging over  $\vp$,  are large: $p^2 \ga M^2\brho$. Hence  the higher order in $\vp$  terms 
in \eq{Sind} may
and will, as we see below, then play an important role. The reason why they survive is, roughly speaking,
a quadratic divergence of the involved integrals.
These terms are however subordinated in $h$ because 
$
\LA \vp(-p) \vp(p) \RA \propto h
$
owing to \eq{rhorho}. 

It is instructive to give yet another explanation why the higher terms can emerge. Let us consider $T_{zz}$ given by 
Eqs.~\rf{Tzz} and average  \rf{Tzzmat} over the regulator fields  with $\vp$ playing again the role of an
external field. The result is given by the diagrams in Fig.~\ref{mean-F1} whose analytic expressions are listed in
Eqs.~\rf{A6} -- \rf{A10} of Appedix~\ref{appA}, where it is explicitly shown the cancellation of higher than quadratic
terms when all momenta squared of external lines 
are small (\ie $p_i^2\ll M^2 \brho$).
 I do not see again 
any reason to expect such a cancellation for momenta of the order of $ M^2 \brho$, so counterparts of the higher terms in 
\eq{Sind} may emerge. 

\begin{figure}
\includegraphics[width=12cm]{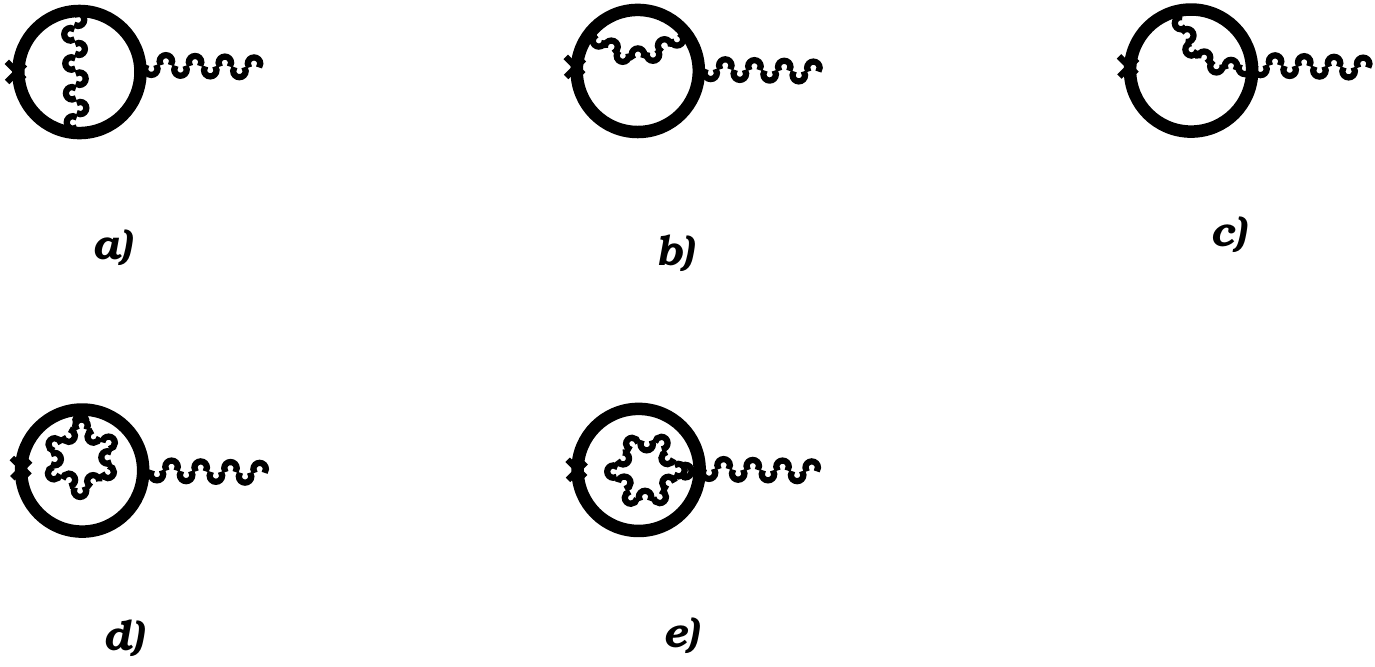} 
\caption{Diagrams associated with the next order correction  to the mean field  for $T_{zz}$.
The combinatorial factors are $a)+2b)-2c)-d)+\half e)$.}
\label{semi}
\end{figure}

The result of the averaging over the regulators  will not be yet the energy-momentum
(pseudo)tensor because the averages in the path-integral language are associated with $T$-products in the operator
language. To obtain a genuine $T_{zz}$, we have to normal order the operators $\bvp$ which produces additional terms 
like the diagrams in Fig.~\ref{semi} 
coming from normal ordering in $\bvp^4$. We thus write 
\be
{\boldsymbol T}^\vp_{zz}= \frac{1}{2} \left(
\frac 1{2b^2}\!:\!\partial_z \bvp\, \partial_z \bvp\!: - \,Q\partial^2_z \bvp\right)+{\cal O}(\bvp^3)
\label{bTzz}
\ee
again in the spirit of DDK.

One more source of the nonlinearity is the well-known fact that the norm of $\vp$ is nonlinear
\be
||\delta \vp||^2 = \int \d^2 z \rho(z) (\delta \vp(z))^2.
\label{nonli}
\ee
We can adopt the philosophy of DDK and replace the path integral over $\vp$ with  the nonlinear norm \rf{nonli}  by 
 the path integral over the field $\vp_0$ with a linear one
\be
||\delta \vp_0||^2 = \int \d^2 z \brho (\delta \vp_0(z))^2 
\ee
by introducing the Jacobian for the transformation from $\vp$
to $\vp_0$. It has again the form of (the exponential of) the  action \rf{Sind} and simply changes its coefficients. 
We shall therefore replace in \eq{Sind}
\be
\frac{26-d}{6h} \Rightarrow \frac{1}{b_0^2}, \qquad  \frac1{b_0^2}= \frac{26-d} {6h}+{\cal O}(1).
\label{70}
\ee
The difference between this $b_0^2$ and $b^2$ in \eq{Seff} comes to order $h$ from the  diagrams with one propagator
which are computed below.

My last comment before proceeding with the computations is that
the propagator $\LA \vp(-p) \vp(p) \RA$
behaves as $1/p^2$ for small $p^2$, so one might expect therefore logarithmic
IR divergences, associated with this behavior, which would spoil conformal invariance. 
However, the low-momentum effective action is quadratic in the variable
$\vp$ as is already mentioned (and
demonstrated by explicit computations in Appendix~\ref{appA}), so the divergences are expected
to cancel each other because the induced action coincides with the effective action
in the IR domain. We shall see in explicit computations this is indeed the case. The remaining contribution to be 
calculated will come from virtual momenta squared of the order of the cutoff: $k^2 \sim M^2 \brho$.
I believe this is a heuristic proof of the {\em theorem} about the cancellation of the IR divergences.

\subsection{Correction to $\boldsymbol {T_{zz}}$} 

The diagrams of the next to the leading order in $h$ which describe ``quantum''
corrections to the mean-field  approximation for $T_{zz}$ are depicted in Fig.~\ref{semi}.
Their analytic expressions are listed in Eqs.~\rf{B1} -- \rf{B5} of Appendix~\ref{appB}.

Every individual diagram has an IR divergence coming from the $\vp$-$\vp$ propagator, but it 
has indeed canceled in the sum  as anticipated. 
Actually the cancellation happens for the sum $a)+2b)-2c)$ because
$
d)=\half e)
$
so only the diagrams in Fig.~\ref{semi}$a$, \ref{semi}$b$ and \ref{semi}$c$ contribute
resulting in
\be
a)+2b)-2c)=-\frac{13p_z^2}{288}.
\label{333}
\ee

It is instructive to present the result in the DDK form
\be
T_{zz}^\varphi = \frac{1}{2}\left(\frac1{2b^2}\partial_z \vp \partial_z \vp -Q \partial_z^2 \vp \right). 
\label{Tzz0}
\ee
Multiplying \rf{333} by the normalizations of the propagator \rf{rhorho} and of the integrals and summing with the 
leading-order diagrams in Fig.~\ref{mean-F1}, we obtain for the coefficient $Q$ in \eq{bTzz} 
\be
Q= \frac{q_0}{b_0^2} -\frac{13}{6} +{\cal O}(h),
\label{Q1}
\ee
where $b_0^2$ is defined in \eq{70}.

An analogous direct computation of  the quadratic in $\vp$ term in \eq{bTzz} is a bit more tedious and
involves 12 diagrams: 7 of which are new, while the contribution of the sum of remaining 5 diagrams is like
$\half Q \partial^2 \vp^2$. The integrals involve two external momenta, which complicates their computation.

\subsection{Correction to $\boldsymbol {S_{\rm eff}}$}

\begin{figure}
\includegraphics[width=13cm]{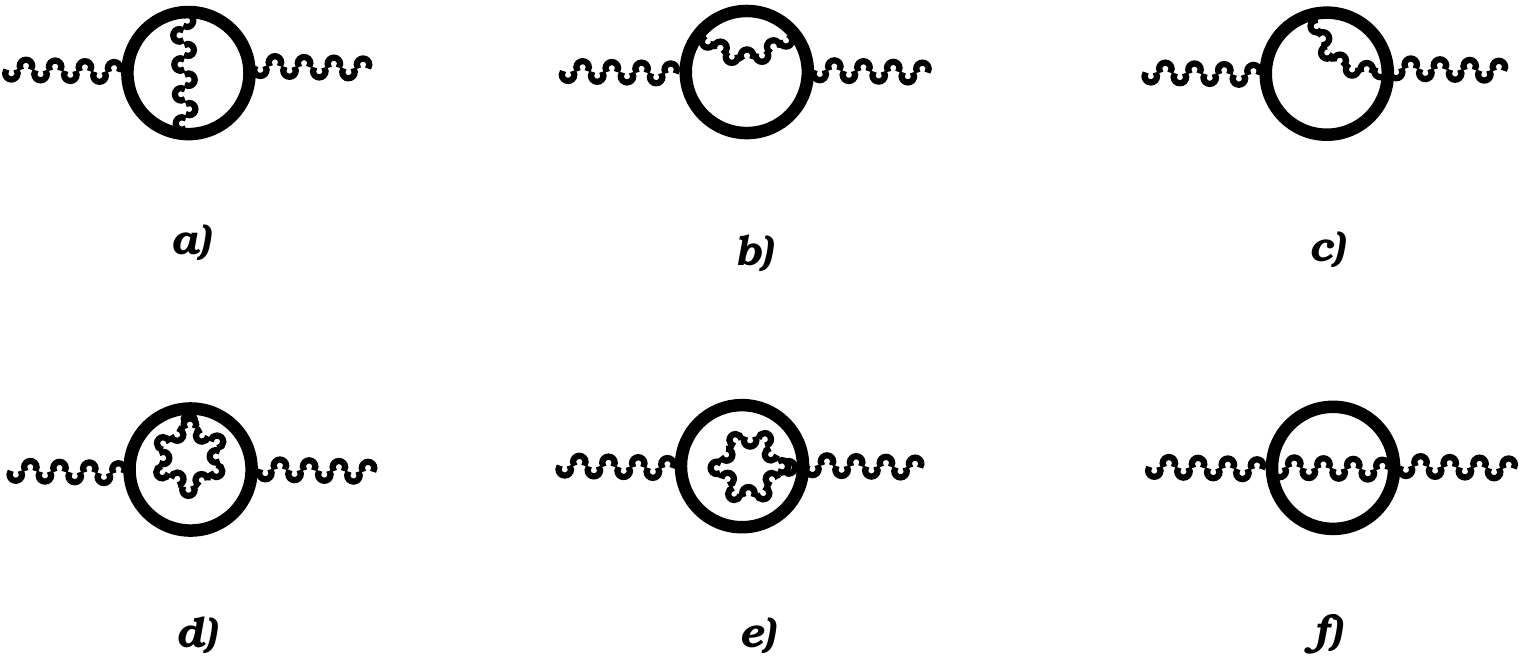} 
\caption{Diagrams associated with the next order correction  to the mean field  for $S_{\rm eff}$.
The combinatorial factors are $a)+2b)-4c)-d)+e)+f)$.  } 
\label{semi-2}
\end{figure}

The diagrams of the next to the leading order in $h$ which describe ``quantum''
corrections to the mean-field  approximation for $S_{\rm eff}$ are depicted in Fig.~\ref{semi-2}.
Their analytic expressions are listed in Eqs.~\rf{B9} -- \rf{B14} of Appendix~\ref{appB}.
Every individual diagram has again an IR divergence coming from the $\vp$-$\vp$ propagator, but it 
has  canceled in the sum  as anticipated. 
Actually the cancellation happens for the sum $a)+2b)-4c)+f)$ because
$
d)= e),
$
so only the diagrams in Fig.~\ref{semi-2}$a$, \ref{semi-2}$b$, \ref{semi-2}$c$ and \ref{semi-2}$f$ contribute.
Accounting for combinatorial factors, 
we obtain 
\be
a)+2b)-4c)+f)=-\frac{5p^2}{48}.
\label{555}
\ee

Multiplying \rf{555} by the normalization of the propagator \rf{rhorho} and of  the integrals,
accounting for ghosts and summing  
with the mean-field result, we obtain for the coupling constant in the effective action~\rf{Seff}
\be
\frac{1}{b^2}=\frac{1}{b^2_0}-5+{\cal O}(h) .
\label{b1}
\ee

\subsection{Remark on the Universality}

In the above computations of $T_{zz}$ and $S_{\rm eff}$ we substituted $G(p)$ in \eq{Sind} by $p^2$ because 
otherwise the computation is hopeless. A question arises as to whether this affects the results because 
characteristic virtual momenta squared in the diagrams are $\sim M^2\brho$.

 It is possible to verify th by
changing the regularization procedure \rf{PV22} to
\be
 \tr\log(-{\cal O})\big|_{\rm reg}= -
\int_{0}^\infty \frac{\d \tau}{\tau} \,\tr \e^{\tau{\cal O}}
\left(1-\e^{-\tau M^2}\right)^N .
\label{PVNN}
\ee
Such a modification of the Pauli-Villars regularization is discussed in \cite{AM17c}.
The regularization \rf{PVNN} involves $N$ Pauli-Villars regulators with masses $\sqrt{n} M$ ($n=1,\ldots,N$) which
complicates the computation. It can be shown however that the results \rf{Q1}, \rf{b1} do not change which is an argument 
in favor of their universality.

\section{Discussion\label{s:discu}}

The main result of this Paper is that a quantization of the effective string about
the mean-field ground state works in $2<d<26$. The mean-field approximation corresponds 
to conformal field theory with the central charge vanishing for any $d$, resulting in
the Alvarez-Arvis ground-state energy 
and complimenting the Polchinski-Strominger approach.
A ``semiclassical'' expansion about the mean field can be treated adopting the philosophy of DDK.

The difference from DDK is that our $\vp$ is massless as a consequence of 
 the minimization at the mean-field saddle point.
 The massless $\vp$ is thus 
a consequence of the nonperturbative mean-field ground state for $2<d<26$ in
contrast to the usual perturbative one for $d<2$.
This
 may lead to infrared logarithms which would spoil conformal 
invariance when accounting for fluctuations about the mean field, but we argued they have to cancel 
because the low-momentum (or effective) action is quadratic in $\vp$. This cancellation is explicitly shown
to the lowest order  of the ``semiclassical'' expansion about the mean field. Thus,we expect 
that conformal invariance should be maintained order by order of the expansion.

The explicit computation shows the (induced) action governing fluctuations
about the mean field is however not
quadratic in $\vp$, while only its low-momentum limit -- the effective action --  is quadratic.
The reason for that is, roughly speaking, quadratic divergences of the involved integrals.
Using the Pauli-Villars regularization, I have shown how to systematically  treat the 
induced action \rf{effgen} and 
to deal with these higher order
 terms
without assuming that $\vp$ is smooth 
and  the determinants are approximated by the conformal anomaly. 
Their emergence 
may influence the results  and deserve further investigation.

The most interesting question is what would be the spectrum of the Nambu-Goto string 
beyond the mean-field approximation. In particular, whether the universal correction
to the Alvarez-Arvis spectrum at the
 order $1/\beta^5$ (see \cite{AK13} and references therein) is reproduced 
in the ``semiclassical'' expansion
about the mean field at one loop. This issue will be considered elsewhere.

\subsection*{Acknowledgment }
I am grateful to Jan Ambj\o rn for sharing his insight into Strings.

\appendix

\section{Leading-order explicit computations\label{appA}}

To the leading order in $h$ we consider $\delta \rho$ (or $\vp$) as an external field over which we shall
average to next orders.

\subsection{$\boldsymbol{T_{a}^a}$}

The contribution of diagrams in Fig.~\ref{mean-F1} to $T_a^a$ from the regulators read
\be
a)=\int \frac{\d^2 k}{(2\pi)^2} \left\{ \frac{2M^2}{(k^2+2M^2)}-\frac{2M^2}{(k^2+M^2]}   \right\}
= -\frac{M^2}{2\pi} \log2,
\ee
\be
b)=\int \frac{\d^2 k}{(2\pi)^2} \left\{ \frac{4M^4}{(k^2+2M^2)[(k-p)^2+2M^2]}-\frac{2M^4}{(k^2+M^2)[(k-p)^2+M^2]}   \right\}
\stackrel{M\to\infty}= \frac{p^2}{24\pi},
\ee
\bea
c)&=&\int \frac{\d^2 k}{(2\pi)^2} \left\{ \frac{8M^6}{(k^2+2M^2)[(k-p)^2+2M^2][(k-p-q)^2+2M^2]} \right.\non&& \left.
-\frac{2M^6}{(k^2+M^2)[(k-p)^2+M^2][(k-p-q)^2+M^2]}   \right\}
\stackrel{M\to\infty}= \frac{p^2+q^2+pq}{24\pi},
\eea
\bea
d)&=&\int \frac{\d^2 k}{(2\pi)^2} \left\{ \frac{16M^8}{(k^2+2M^2)[(k-p)^2+2M^2][(k-p-q)^2+2M^2][(k-p-q-r)^2+2M^2]}
 \right.\non &&\left.
-\frac{2M^8}{(k^2+M^2)[(k-p)^2+M^2][(k-p-q)^2+M^2][(k-p-q-r)^2+M^2]}   \right\} \non &&
\stackrel{M\to\infty}=\frac{3p^2+4q^2+3r^2+4pq+2pr+4qr}{80\pi},
\eea
\bea
e)&=&\int \frac{\d^2 k}{(2\pi)^2} \left\{ \frac{32M^{10}}{(k^2+2M^2)\cdots[(k-p-q-r)^2+2M^2][(k-p-q-r-t)^2+2M^2]}
 \right.\non &&\left.
-\frac{2M^{10}}{(k^2+M^2)[(k-p)^2+M^2]\cdots[(k-p-q-r)^2+M^2][(k-p-q-r-t)^2+M^2]}   \right\} \non && 
\stackrel{M\to\infty}=\frac{2p^2+3q^2+3r^2+2t^2+3pq+2pr+pt+4qr+2qt+3rt}{60\pi}.
\eea

Multiplying each wavy line by $-\delta \rho$, passing to the coordinate space and summing up the 
contributions of the diagrams in Fig.~\ref{mean-F1} with these of ghosts, we find
\bea
\LA T_a^a \RA&=& \frac{26-d}{12h} \left[ \partial^2 \dr -(2 \dr\, \partial^2 \dr +\partial_a \dr\, \partial_a \dr )
+3(\dr^2 \partial^2\dr+\dr \,\partial_a \dr\, \partial_a \dr) \right . \non
&&\left.-(4\dr^3 \partial^2\dr+6\dr^2 \,\partial_a \dr\, \partial_a \dr)   \right]+{\cal O}(\dr^5)\non
&=&  \frac{26-d}{12h} \left( 1-\vp +\frac 12 \vp^2-\frac16 \vp^3\right) \partial^2 \vp+{\cal O}(\vp^5),
\label{A11}
\eea
where we have used \rf{drho}
and expanded in $\vp$. We have thus reproduced \eq{53} to this order.

In \eq{A11} we simply subtracted $26$ from $d$ to account  for the ghost contribution because the
contribution of diagrams which emerge from ghost and the regulators of ghosts are identical for the 
mean-field and perturbative vacua just as it is for the matter fields and regulators.
The same applies below for $T_{zz}$.

\subsection{$\boldsymbol {T_{zz}}$}

The analogous contribution of diagrams in Fig.~\ref{mean-F1} to $T_{zz}$ read
\be
a)=\int \frac{\d^2 k}{(2\pi)^2} \left\{ \frac{k_z^2}{(k^2+2M^2)}-\frac{ k_z^2}{(k^2+M^2]}   \right\}
=0,
\label{A6}
\ee
\be
b)=\int \frac{\d^2 k}{(2\pi)^2} \left\{ \frac{2M^2k_z(k_z-p_z)}{(k^2+2M^2)[(k-p)^2+2M^2]}-\frac{2M^2k_z(k_z-p_z)}{(k^2+M^2)[(k-p)^2+M^2]}   \right\}
\stackrel{M\to\infty}= \frac{p^2_z}{24\pi},
\ee
\bea
c)&=&\int \frac{\d^2 k}{(2\pi)^2} \left\{ \frac{4M^4k_z(k_z-p_z-q_z)}{(k^2+2M^2)[(k-p)^2+2M^2][(k-p-q)^2+2M^2]}
 \right.\non&& \left.
-\frac{2M^4k_z(k_z-p_z-q_z)}{(k^2+M^2)[(k-p)^2+M^2][(k-p-q)^2+M^2]}   \right\}
\stackrel{M\to\infty}= \frac{p^2_z+q^2_z+3p_zq_z}{48\pi},
\eea
\bea
d)&=&\int \frac{\d^2 k}{(2\pi)^2} \left\{ \frac{8M^6k_z(k_z-p_z-q_z-r_z)}{(k^2+2M^2)[(k-p)^2+2M^2][(k-p-q)^2+2M^2][(k-p-q-r)^2+2M^2]}
 \right.\non &&\left.
-\frac{2M^6k_z(k_z-p_z-q_z-r_z)}{(k^2+M^2)[(k-p)^2+M^2][(k-p-q)^2+M^2][(k-p-q-r)^2+M^2]}   \right\} \non && 
\stackrel{M\to\infty}=\frac{3p_z^2+4q_z^2+3r_z^2+9p_zq_z+12p_zr_z+9q_zr_z}{240\pi},
\eea
\bea
e)&=&\int \frac{\d^2 k}{(2\pi)^2} \left\{ \frac{16M^{8}k_z(k_z-p_z-q_z-r_z-t_z)}
{(k^2+2M^2)\cdots[(k-p-q-r)^2+2M^2][(k-p-q-r-t)^2+2M^2]}
 \right.\non &&\left.
-\frac{2M^{8}k_z(k_z-p_z-q_z-r_z-t_z)}{(k^2+M^2)[(k-p)^2+M^2]\cdots[(k-p-q-r)^2+M^2][(k-p-q-r-t)^2+M^2]}   \right\} \non && 
\stackrel{M\to\infty}=\frac{2p^2_z+3q^2_z+3r^2_z+2t^2_z+6p_zq_z+8p_zr_z+10p_zt_z
+7q_zr_z+8q_zt_z+6r_zt_z}{240\pi}.
\label{A10}
\eea

Using  \rf{drho}, we analogously to \eq{A11}
obtain for $T^\vp_{zz}$ 
\bea
\LA T_{zz}\RA&=&\frac{26-d}{12h}\left[ \partial^2_z \dr -( \dr\, \partial^2_z \dr +\frac32\partial_z \dr\, \partial_z \dr )
+(\dr^2 \partial_z^2\dr+3\dr \,\partial_z \dr\, \partial_z \dr) \right . \non
&&\left.-(\dr^3 \partial^2_z\dr+\frac92\dr^2 \,\partial_z \dr\, \partial_z \dr)   \right]+{\cal O}(\dr^5)\non
&=&  \frac{26-d}{12h} \left( \partial^2_z \vp-\frac 12\partial_z \vp\, \partial_z \vp \right)+{\cal O}(\vp^5),
\eea
\ie the free energy-momentum tensor to this order.

The reason why I presented in this Appendix
the explicit computations of $T^{a}_a$ and $T_{zz}$ is to emphasize that numerical factors are most
important to get the free-theory results. The cancellation would no longer take place if these factors were 
changed due to induced interactions, as we shall immediately see in the next Appendix.

\section{``Semiclassical'' corrections\label{appB}}

\subsection{Contribution to $\boldsymbol{T_{zz}}$ }

The contributions of diagrams in Fig.~\ref{semi} to $T_{zz}$ involve
\bea
a) = \int \frac{\d^2 k \d^2q}{(2\pi)^2} 
 \left\{ \frac{8M^6k_z(k_z-p_z)}{(k-q)^2[(k-p)^2+2M^2][(q-p)^2+2M^2](k^2+2M^2)(q^2+2M^2)} \right. \non \left.-
\frac{2M^6k_z(k_z-p_z)}{(k-q)^2[(k-p)^2+M^2][(q-p)^2+M^2](k^2+M^2)(q^2+M^2)}   \right\},~~~ 
\label{B1}
\eea
\bea
b) = \int \frac{\d^2 k \d^2q}{(2\pi)^2} 
 \left\{ \frac{8M^6k_z(k_z-p_z)}{(k-q)^2[(k-p)^2+2M^2](k^2+2M^2)^2(q^2+2M^2)} \right. \non \left.-
\frac{2M^6k_z(k_z-p_z)}{(k-q)^2[(k-p)^2+M^2](k^2+M^2)^2(q^2+M^2)}   \right\} ,
\eea
\bea
c) = \int \frac{\d^2 k \d^2q}{(2\pi)^2} 
 \left\{ \frac{4M^4k_z(k_z-p_z)}{(k-q)^2[(k-p)^2+2M^2](k^2+2M^2)(q^2+2M^2)}\right. \non \left.
- \frac{2M^4k_z(k_z-p_z)}{(k-q)^2[(k-p)^2+M^2](k^2+M^2)(q^2+M^2)}   \right\} ,
\eea
\be
d) = \int \frac{\d^2 k \d^2q}{(2\pi)^2} 
 \left\{ \frac{4M^4k_z(k_z-p_z)}{q^2[(k-p)^2+2M^2](k^2+2M^2)^2} 
-\frac{2M^4k_z(k_z-p_z)}{q^2[(k-p)^2+M^2](k^2+M^2)^2}   \right\} ,
\ee
\be
e) = \int \frac{\d^2 k \d^2q}{(2\pi)^2} 
 \left\{ \frac{2M^2k_z(k_z-p_z)}{q^2[(k-p)^2+2M^2](k^2+2M^2)} 
-\frac{2M^2k_z(k_z-p_z)}{q^2[(k-p)^2+M^2](k^2+M^2)}   \right\} ,
\label{B5}
\ee

For the computation of integrals it is convenient to multiply a generic integral 
\be
\int \frac{\d^2 k}{(2\pi)^2} k_a (k_b-p_b) f(k^2,p^2,kp) = f_1(p^2) g_{ab} +f_2 (p^2) p_a p_b
\ee
by the projector 
\be
P^{ab}=2\frac{p^a p^b}{p^2}-g^{ab}
\ee
to get 
\be
\int \frac{\d^2 k}{(2\pi)^2} \left(2\frac{(kp)^2}{p^2}-kp -k^2 \right) f(k^2,p^2,kp) =f_2 (p^2) p^2.
\ee
Then we have
\be
\int \frac{\d^2 k}{(2\pi)^2} k_z (k_z-p_z) f(k^2,p^2,kp) = f_2 (p^2) p_z^2 .
\ee
This trick is implemented in the Mathematica program of  Appendix~\ref{appM}, 
where the integrals are computed by
first integrating over the two relative angles and then by the two absolute values of the virtual momenta,

Performing the computation by the Mathematica program in Appendix~\ref{appM} 
and accounting for combinatorial factors, 
we obtain 
\be
a)+2b)-2c)-d)+\half e)=-\frac{13p_z^2}{288}.
\label{333a}
\ee
Notice that 
$ 
d)=\half e)
$ 
so only the diagrams in Fig.~\ref{semi}$a$, \ref{semi}$b$ and \ref{semi}$c$ contribute.
The infrared divergence coming from the $\vp$-$\vp$ propagator has indeed canceled in the sum,
as anticipated.

Multiplying \rf{333a} by the normalization of the propagator \rf{rhorho} and of the integrals 
and summing the diagrams 
in Fig.~\ref{mean-F1} and Fig.~\ref{semi}, we obtain
\be
T^\vp_{zz}= \frac{26-d}{12h}\left[ \left( 1-\frac{13h}{26-d}\right)\partial^2_z \vp-\frac 12\partial_z \vp\, \partial_z \vp \right]
+{\cal O}(h).
\ee

\subsection{Contribution to $\boldsymbol{S_{\rm eff}}$ }

The contributions of diagrams in Fig.~\ref{semi-2} to $S_{\rm eff}$ involve
\bea
a) = \int \frac{\d^2 k \d^2q}{(2\pi)^2} 
 \left\{ \frac{16M^8}{(k-q)^2[(k-p)^2+2M^2][(q-p)^2+2M^2](k^2+2M^2)(q^2+2M^2)} \right. \non \left.-
\frac{2M^8}{(k-q)^2[(k-p)^2+M^2][(q-p)^2+M^2](k^2+M^2)(q^2+M^2)}   \right\}
 ,~~~~~~
\label{B9}
\eea
\bea
b) = \int \frac{\d^2 k \d^2q}{(2\pi)^2} 
 \left\{ \frac{16M^8}{(k-q)^2[(k-p)^2+2M^2](k^2+2M^2)^2(q^2+2M^2)} \right. \non \left.-
\frac{2M^8}{(k-q)^2[(k-p)^2+M^2](k^2+M^2)^2(q^2+M^2)}   \right\} ,
\eea
\bea
c) = \int \frac{\d^2 k \d^2q}{(2\pi)^2} 
 \left\{ \frac{8M^6}{(k-q)^2[(k-p)^2+2M^2](k^2+2M^2)(q^2+2M^2)}\right. \non \left.
- \frac{2M^6}{(k-q)^2[(k-p)^2+M^2](k^2+M^2)(q^2+M^2)}   \right\} ,
\eea
\be
d) = \int \frac{\d^2 k \d^2q}{(2\pi)^2} 
 \left\{ \frac{8M^6}{q^2[(k-p)^2+2M^2](k^2+2M^2)^2} 
-\frac{2M^6}{q^2[(k-p)^2+M^2](k^2+M^2)^2}   \right\} ,
\ee
\be
e) = \int \frac{\d^2 k \d^2q}{(2\pi)^2} 
 \left\{ \frac{4M^4}{q^2[(k-p)^2+2M^2](k^2+2M^2)} 
-\frac{2M^4}{q^2[(k-p)^2+M^2](k^2+M^2)}   \right\} ,
\ee
\bea
f) &= &\int \frac{\d^2 k \d^2q}{(2\pi)^2} 
 \left\{ \frac{4M^4}{(k-q)^2[(k-p)^2+2M^2](q^2+2M^2)} \right. \non 
&&\hspace*{2cm}\left.
-\frac{2M^4}{(k-q)^2[(k-p)^2+M^2](q^2+M^2)}   \right\} .
\label{B14}
\eea

Performing the computation by the Mathematica program of Appendix~\ref{appM} 
and accounting for combinatorial factors, 
we obtain 
\be
a)+2b)-4c)-d)+ e)+f)=-\frac{5p^2}{48}.
\label{555a}
\ee
Notice that again
$ 
d)= e)
$ 
so only the diagrams in Fig.~\ref{semi-2}$a$, \ref{semi-2}$b$, \ref{semi-2}$c$ and \ref{semi-2}$f$ contribute.
The infrared divergence coming from the $\vp$-$\vp$ propagator has indeed canceled in the sum,
as anticipated.

Multiplying \rf{555a} by the normalization of the propagator \rf{rhorho} and of the integrals, 
accounting for ghosts and summing  
with the mean-field result, we obtain
\be
S_{\rm eff}= \frac{26-d}{96\pi h}\left( 1-\frac{30h}{26-d}\right) \int \d^2 z\,
\partial_a \vp \partial_a \vp
+{\cal O}(h).
\ee

\section{Mathematica programs\label{appM}}

This Appendix can be downloaded as an ancillary file \cite{appC}.


\end{document}